\documentclass[%
reprint,
superscriptaddress,
preprintnumbers,
amsmath,amssymb,
aps,
prd,
]{revtex4-2}

\usepackage{float}
\usepackage{graphicx}
\usepackage{dcolumn}
\usepackage{bm}
\usepackage{bbold}
\usepackage{braket}
\usepackage{amssymb,amsmath}
\usepackage{bbm}

\usepackage{color}
\usepackage{amsfonts}
\usepackage{subfigure}
\usepackage{array}
\usepackage{enumerate}

\newcolumntype{L}{>{\centering\arraybackslash}m{3cm}}

\definecolor{blue}{rgb}{0,0,1}

\definecolor{green}{rgb}{0,1,0}

\definecolor{red}{rgb}{1,0,0}

\definecolor{gray}{rgb}{.5,.5,.5}

\definecolor{darkgreen}{rgb}{.0,.5,.0}

\usepackage{xspace}
\usepackage{siunitx}
\usepackage{xfrac}
\usepackage{hyperref}
\usepackage[nameinlink]{cleveref}
\usepackage{appendix}
\usepackage{xifthen}
\usepackage{xcolor}
\hypersetup{
	colorlinks,
	linkcolor={red!75!black},
	citecolor={blue!75!black},
	urlcolor={blue!75!black}
}
\def\Fig#1{\Cref{#1}}

\def\Eq#1{\Cref{#1}}

\def\eqref#1{\Cref{#1}}

\def\sec#1{\Cref{#1}}
\def\app#1{\hyperref[#1]{App.~\ref{#1}}}
\def\app#1{\Cref{#1}}

\def\lA0{{\langle A_0 \rangle}}
\def\bA0{{\bar{A}_0}}

\def\0#1#2{\frac{#1}{#2}}

\graphicspath{{./figures/}{./}}

\begin{document}
	
\title{Estimates on the convergence of expansions at finite baryon chemical potentials}

\author{Rui Wen}
\email{rwen@ucas.ac.cn}
\affiliation{School of Nuclear Science and Technology, University of Chinese Academy of Sciences, Beijing, 100049,  P.R. China}
  
\author{Shi Yin}
\affiliation{School of Physics, Dalian University of Technology, Dalian, 116024,
  P.R. China}
	
\author{Wei-jie Fu}
\affiliation{School of Physics, Dalian University of Technology, Dalian, 116024,
  P.R. China}
	
	
\begin{abstract}

Convergence of three different expansion schemes at finite baryon chemical potentials, including the conventional Taylor expansion, the Pad\'e approximants, and the $T'$ expansion proposed recently in lattice QCD simulations, have been investigated in a low energy effective theory within the fRG approach. It is found that the $T'$ expansion or the Pad\'e approximants would hardly improve the convergence of expansion in comparison to the conventional Taylor expansion, within the expansion orders considered in this work. Furthermore, we find that the consistent regions of the three different expansions are in agreement with the convergence radius of the Lee-Yang edge singularities.

\end{abstract}

\maketitle
	
\section{Introduction}
\label{sec:int}

Knowledge of the QCD phase structure, and in particular locating the critical end point (CEP), has been widely discussed during the past few decades. Both experimental and theoretical studies have made significant progress. On the experimental side, the Beam Energy Scan (BES) Program at the Relativistic Heavy Ion Collider (RHIC) have measured the high order cumulants of net-proton distributions \cite{STAR:2020tga, STAR:2021iop, STAR:2021rls, STAR:2021fge, Pandav:2020uzx}, which are thought of as ideal probe to detect the CEP. On the theoretical side, lattice QCD simulations are employed to calculate the equation of state \cite{Bazavov:2009zn, Bazavov:2014pvz, Borsanyi:2013bia} and baryon number fluctuations \cite{HotQCD:2017qwq, Borsanyi:2018grb, Bazavov:2020bjn, Borsanyi:2023wno, Bollweg:2021vqf} at vanishing chemical potentials. However, lattice simulations are limited to imaginary and vanishing chemical potential due to the sign problem. On the other hand, the first principles functional approaches, such as the functional renormalization group (fRG) \cite{Fu:2019hdw, Braun:2020ada, Mitter:2014wpa, Cyrol:2017ewj, Braun:2014ata} and Dyson-Schwinger equation (DSE) \cite{Gao:2020fbl, Bernhardt:2021iql, Isserstedt:2019pgx, Lu:2023mkn}, do not suffer from the sign problem and allow us to do direct calculations at real chemical potentials. Recently, the functional approaches have shown convergent estimates for the location of CEP in the phase diagram, which is in a small region of  $ \mu_B \in (600,650) $ MeV with $\mu_B/T>4$ \cite{Fu:2019hdw, Gao:2020fbl, Isserstedt:2019pgx}. In the region of large baryon chemical potentials, errors of calculation resulting from truncations in the functional methods might increase sizably, which have to be controlled carefully via, e.g., comparison with lattice QCD and functional QCD of improved truncations. For recent reviews, see, e.g.  \cite{Guenther:2022wcr, Guenther:2020jwe, Dupuis:2020fhh, Rennecke:2019lus, Fu:2022gou}.

In the past few years, lots of efforts have been made to extend lattice QCD results into finite chemical potentials. In \cite{Bazavov:2017dus, HotQCD:2017qwq}, Taylor expansion method is employed to extend the equation of state and the baryon number fluctuations into non-zero chemical potential. Due to the negative values of both $\chi_6^B$ and higher order susceptibilities \cite{Borsanyi:2018grb, Bazavov:2020bjn, Borsanyi:2023wno}, the pressure and baryon-number density with the Taylor expansion method show non-monotonic behavior at $\mu_B/T \gtrsim 2.5$ \cite{Bazavov:2017dus}, which implies the Taylor expansion method becomes less reliable in this region. It is generally believed that the convergence radius of the Taylor expansion is limited by singularities in the complex plane of chemical potentials, such as the Lee-Yang edge singularities \cite{Mukherjee:2019eou, Borsanyi:2013hza, Basar:2021gyi, Connelly:2020gwa, Mukherjee:2021tyg, Karsch:2023rfb} and the Roberge-Weiss transition singularities \cite{Vovchenko:2017gkg}. Besides, the Pad\'e resummation \cite{Karsch:2010hm, Datta:2016ukp, Pasztor:2020dur, Bollweg:2022rps}  and other resummation method \cite{Mondal:2021jxk, Mukherjee:2021tyg} are also investigated widely, and a sign reweighting method, which allows for direct simulations at real baryon densities, is also proposed in \cite{Borsanyi:2021hbk, Pasztor:2021ray, Borsanyi:2022soo}.

In \cite{Borsanyi:2021sxv}, the Wuppertal-Budapest collaboration has proposed a $T'$ expansion scheme, which extrapolates the pressure and baryon number density at imaginary and vanishing chemical potentials to those at real ones. The rescaling coefficients are related with Taylor expansion coefficients. So in other words, they also provide a resummation scheme of the Taylor expansion coefficients. Since the statistical errors of $\chi_1^B,\chi_2^B$ at imaginary chemical potentials are quite small in lattice QCD simulations, the statistical errors of the extrapolation are well controlled. Recently, the expansion scheme was generalized with the strangeness neutrality condition \cite{Borsanyi:2022qlh}. 

Nonetheless, the convergence radius of the $T'$ expansion scheme is still unclear. In this work, we investigate the convergence radius of the new expansion scheme, and compare it with the Taylor expansion and the Pad\'e resummation. The Polyakov-loop extended quark-meson (PQM) model is employed, which can well describe the chiral symmetry breaking/restoration and confinement/deconfinement phase transitions. Baryon number fluctuations obtained in this effective theory are in good agreement with lattice QCD results \cite{Fu:2015naa, Fu:2021oaw, Fu:2023lcm}.

This paper is organized as follows: In \sec{sec:LEFT}, the Polyakov-loop extended quark-meson model within the fRG approach at real and imaginary chemical potential is introduced. In \sec{sec:Expansion_Scheme}, we briefly review the expansion scheme and give the formulae of higher-order generalized susceptibilities. In \sec{sec:res}, we give the numerical results. The summary and conclusion are given in \sec{sec:sum}.

\section{Two-flavor Polyakov-loop extended quark-meson model}
\label{sec:LEFT}

We employ the two-flavor Polyakov-loop extended quark-meson model within the functional renormalization group approach \cite{Fu:2015naa, Fu:2021oaw, Sun:2018ozp, Fu:2016tey, Fu:2023lcm, Yin:2019ebz, Pawlowski:2014zaa}, which is a QCD low energy effective theory. The Euclidean effective action reads
\begin{align}
\Gamma_k=&\int_x \bigg\{Z_{q,k}\bar{q} \Big [\gamma_\mu \partial_\mu -\gamma_0(\mu+igA_0) \Big ]q+\frac{1}{2}Z_{\phi,k}(\partial_\mu \phi)^2 \nonumber \\
+&h_k\bar{q}\big(T^0\sigma+i\gamma_5\vec{T}\cdot \vec{\pi}\big)q+V_k(\rho)-c\sigma +V_{\mathrm{glue}}(L,\bar L) \bigg\}\,,
\label{eq:action}
\end{align}
with a 4-dimensional integral $\int_x=\int^{1/T}_0dx_0\int d^3x$. Here $T$, $\mu$ are the temperature and the quark chemical potential, respectively. The baryon chemical potential reads $\mu_B=3\mu$. The meson field reads $\phi=(\sigma,\vec{\pi})$ and $V_k(\rho)$ denotes a chiral symmetric effective potential with $\rho=\phi^2/2$. The chiral symmetry is explicitly broken by the linear sigma term $-c \sigma$. $A_0$ denotes the gluon background field. $L(A_0)$  is the traced Polyakov loop and $\bar L(A_0)$ is its conjugate, which are defined as
\begin{align}
L(x)=\frac{1}{N_c}\langle \text{Tr} \mathcal P(x) \rangle,
\quad
\bar L(x)= \frac{1}{N_c}\langle \text{Tr} \mathcal P^\dagger(x) \rangle\,,
\end{align}
with
\begin{align}
\mathcal P(x) = \mathcal P \exp\Big(ig \int _0^\beta d \tau A_0(x,\tau)\Big )\,,
\end{align}
and $\mathcal P$ is the path ordering operator. The $V_{\mathrm{glue}}$ denotes the glue potential, which is a function of the traced Polyakov loop $L$ and $\bar{L}$.

In this work, we adopt the local potential approximation (LPA), where the dependence of the wave function renormalizations and Yukawa coupling on the renormalization group (RG) scale is neglected, i.e. $Z_{q/\phi,k}=1,\partial_k h_k=0$. We choose the ultraviolet cutoff scale $k_{\mathrm{UV}}=700\,\text{MeV}$, the initial potential $V_{\mathrm{UV}}(\rho)=\lambda_1 \rho+ \lambda_2\rho^2/2$ with $\lambda_1=482^2\,\text{MeV}^2, \lambda_2=5.7$. Moreover, one has the Yukawa coupling $h=6.5$ and the explicitly chiral symmetry breaking coefficient $c=1.7 \times 10^6 \,\text{MeV}^3$, fixed by fitting the physical observables in the vacuum: $f_\pi=92\,\text{MeV}$, $m_q=300\,\text{MeV}$, $m_\pi=135\,\text{MeV}$ and $m_\sigma=500\,\text{MeV}$. The parameterization of the Polyakov loop potential is used as same as that in \cite{Fu:2021oaw}.

We proceed with the flow equation of $V_k(\rho)$, which reads
\begin{align}
\partial_k V_k(\rho)=&\frac{k^3}{4 \pi^2}\Big[3 l_0^{(B)}(m_\pi) + l_0^{(B)}(m_\sigma) - 4 N_c N_f l_0^{(F)}(m_f)\Big]
\end{align}
Here, $N_c=3$, $N_f=2$, and $l_0^{(B/F)}$ are bosonic/fermionic loop functions, which can be found in \cite{Fu:2015naa, Fu:2021oaw, Wen:2018nkn}. Note that when the chemical potential is purely imaginary, the fermionic distribution functions are complex-valued. The anti-fermionic distribution function is the complex conjugate of the fermionic one, and thus we are left with a real fermionic loop function, 
\begin{align}
l_0^{(F)}(m_f)&=\frac{k}{3 E}(1-n_f(L,\bar L)-\bar n_f(\bar L, L)) \nonumber \\
&=\frac{k}{3 E}[1-2 \text{Re}(n_f)]\,, 
\end{align}
which ensures that the effective potential and physical observables are real-valued. Whereas, if the chemical potential is a general complex number, the imaginary part of the fermionic loop function would be nonzero.

The chiral pseudo-critical temperature of PQM model is $T_c=215\, \mathrm{MeV}$. Following \cite{Fu:2021oaw}, we use the scale-matching between the $N_f=2$ PQM model and the $N_f=2+1$ QCD:
\begin{align}
T^{(N_f=2+1)}_{\mathrm{QCD}}&=c\,\,T^{(N_f=2)}_{\mathrm{PQM}}\\
{\mu_B}^{(N_f=2+1)}_{\mathrm{QCD}}&=c\,\,{\mu_B}^{(N_f=2)}_{\mathrm{PQM}}\,,
\end{align}
with
\begin{align}
c=\frac{{T_c}^{(N_f=2+1)}_{\mathrm{QCD}}}{{T_c}^{(N_f=2)}_{\mathrm{PQM}}}=0.726 \,.
\end{align}
The critical end point is located at $T_{\mathrm{CEP}}=40\,\mathrm{MeV}$, $\mu_{B_{\mathrm{CEP}}}=667\,\mathrm{MeV}$ after the rescaling.
	
\section{Expansion Scheme}
\label{sec:Expansion_Scheme}

The generalized susceptibilities of the baryon number are defined as $i$-th order derivatives of the normalized pressure:
\begin{align}
\chi_i^B=\frac{\partial^i}{{\partial \hat \mu_B }^i}\frac{p}{T^4}\,,
\end{align}
with $\hat \mu_B=\mu_B/T$. We numerically calculate the generalized susceptibilities up to 8-th order in this work, and an algorithmic differentiation technique for higher-order derivatives calculation is proposed in \cite{Wagner:2009pm}.

The Taylor expansion of the pressure is given as:
\begin{align}
\frac{p(T,\hat \mu_B)-p(T,0)}{T^4}=\sum_{n=1}\frac{1}{(2n)!}\chi^B_{2n}(T,0)  \hat \mu_B^{2n}\,.
\label{eq:Taylor}
\end{align}

In the following, we will introduce another two expansion schemes, i.e., the Pad\'e approximants and $T'$ expansion.

\subsection{Pad\'e approximants}

We introduce the Pad\'e approximants to reconstruct the pressure as a function of $\hat\mu_B^2$:
\begin{align}
P[m,n]&\equiv \frac{p(T,\hat \mu_B)-p(T,0)}{T^4} \nonumber\\
&=\frac{\sum_{i=1}^{n/2} a_i\cdot \hat\mu_B^{2i}}{1+ \sum_{j=1}^{m/2} b_j \cdot \hat\mu_B^{2j}}\,.
\end{align}
Here, the coefficients $a_i,b_i$ are determined by solving the equations: 
\begin{align}
\frac{\partial^i P[m,n]}{{\partial \hat \mu_B} ^i}=\chi_i^B \, .
\end{align}
In this work, we mainly consider $P[4,2]$ and $P[4,4]$, which correspond to reconstructing the first $6$-th and $8$-th order susceptibilities respectively:
\begin{align}
P[4,2]=&\frac{60 \chi_2^B \chi_4^B \hat\mu_B^2 + \big(5 (\chi_4^B)^2  - 2 \chi_2^B \chi_6^B\big)\mu_B^4}{120 \chi_4^B - 4 \chi_6^B \hat\mu_B^2}\,\label{eq:PD42},\\[2ex]
P[4,4]=& \Big[2520 \chi_2^B \big(-5 (\chi_4^B)^2 + 2 \chi_2^B \chi_6^B\big) \hat\mu_B^2 - 30 \big(35 (\chi_4^B)^3\nonumber\\[1ex]
& - 28 \chi_2^B \chi_4^B \chi_6^B + 3 (\chi_2^B)^2 \chi_8^B\big) \mu_B^4\Big] \Big[5040 \nonumber \\[1ex]
&\times\big(-5 (\chi_4^B)^2 + 2 \chi_2^B \chi_6^B \big)  + 60 \big(14 \chi_4^B \chi_6^B \nonumber\\[1ex]
&- 3 \chi_2^B \chi_8^B\big) \hat\mu_B^2 + \big(-28 (\chi_6^B)^2 + 15 \chi_4^B \chi_8^B\big) \hat\mu_B^4\Big]^{-1}\,,\label{eq:PD44}
\end{align}
where the expression of $P[4,4]$ is also shown in \cite{Bollweg:2022rps}. The generalized susceptibilities are $i$-th order derivatives of the Pad\'e approximants of the pressure with respect to $\hat\mu_B$. One can also calculate the Pad\'e approximants of $i$-th order susceptibilities, but we would not do it, since they are not our main concerns in this work. Moreover, the poles of the Pad\'e approximants can be used to estimate the convergence radius of Taylor expansion, e.g., the poles of $P[n,2]$ and $P[n,4]$ are related to the ratio estimator and Mercer-Roberts estimator, respectively \cite{Bollweg:2022rps, Vovchenko:2017gkg}.

\subsection{$T'$ expansion scheme}
\label{subsec:T_exp}
%
\begin{figure}[t]
\includegraphics[width=0.5\textwidth]{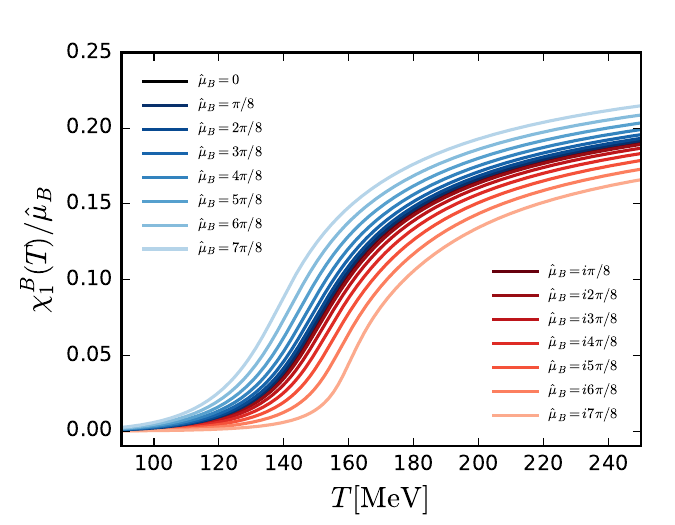}
\caption{Ratio $\chi^B_1/\hat \mu_B$ as a function of the temperature with real (blue) and imaginary (red) baryon chemical potentials, which coincides with the quadratic fluctuation $\chi^B_2$ exactly at $\hat \mu_B=0$. }\label{fig:chi1_muB}
\end{figure}
%

%
\begin{figure}[t]
\includegraphics[width=0.5\textwidth]{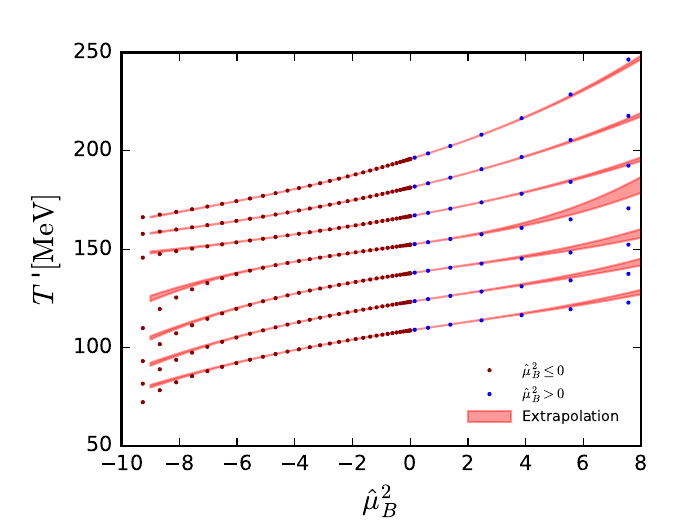}
\caption{Rescaled temperature $T'$ defined in \Eq{eq:chi1_res} as a function of $\hat \mu_B^2$. The original temperature $T$ is chosen in the range of $[108, 196]$ MeV at the interval of 14.5 MeV. The dark-red dots and blue dots stand for results at imaginary and real chemical potentials, respectively. The red bands denote the extrapolation of \Eq{eq:T_res} from the imaginary to real chemical potentials.}\label{fig:con_ext}
\end{figure}
%

%
\begin{figure}[t]
\includegraphics[width=0.5\textwidth]{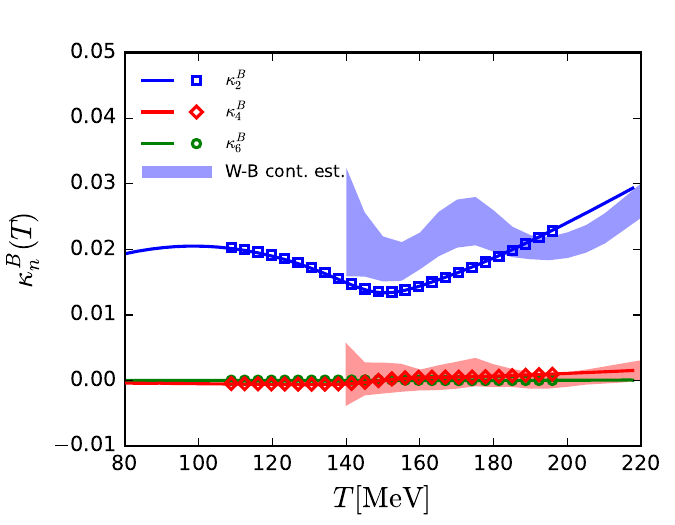}
\caption{Coefficients in \Eq{eq:T_res} as functions of the temperature. The markers denote the results obtained from the fitting of imaginary chemical potentials, and the solid lines stand for those calculated Taylor expansion coefficients in Equations \labelcref{eq:Talyorcalkappa2} through \labelcref{eq:Talyorcalkappa6}. The bands indicate lattice results in \cite{Borsanyi:2021sxv}.}\label{fig:kappa24}
\end{figure}
%

%
\begin{figure*}[t]
\includegraphics[width=0.84\textwidth]{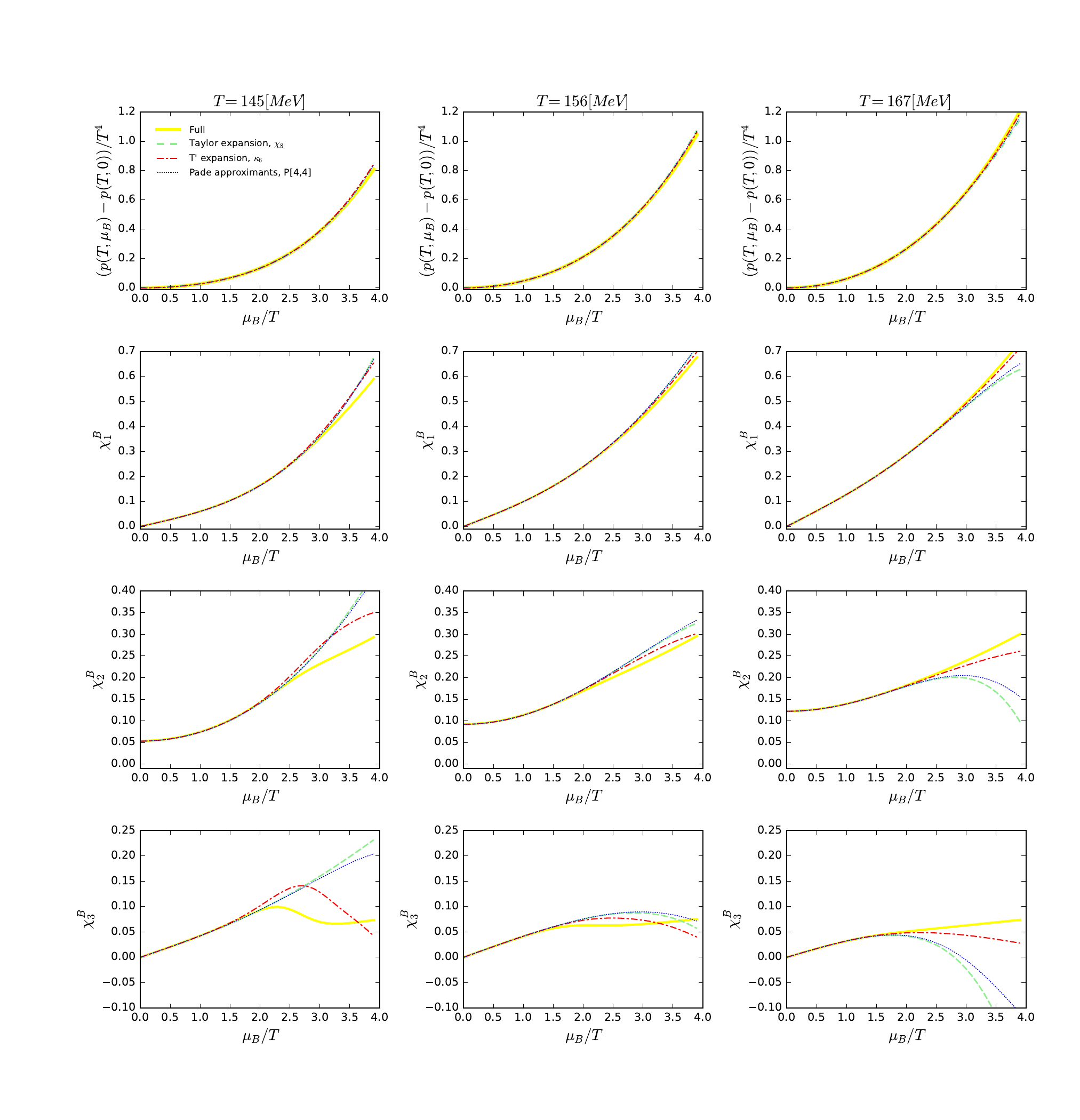}
\caption{Comparison between the direct calculations of the pressure and the first three order generalized susceptibilities and those with the Taylor expansion, Pad\'e approximants and the $T'$ expansion, as functions of the baryon chemical potential divided by the temperature $\mu_B/T$ at three different values of temperature (different columns).}\label{fig:pre_sus}
\end{figure*}
%

Recently, a relation between the baryon number density $n_B(\hat \mu_B)$ at finite chemical potentials and the quadratic baryon number fluctuation $\chi_2^B(\hat\mu_B=0)$ at vanishing chemical potentials has been proposed by the Wuppertal-Budapest collaboration \cite{Borsanyi:2021sxv}, which reads
\begin{align}
\frac{\chi_1^B(T,\hat \mu_B)}{\hat \mu_B}=\chi_2^B(T',0)\,,
\label{eq:chi1_res}
\end{align}
with
\begin{align}
T'= & T\Big(1+\kappa_2^B(T) \hat \mu_B^2+\kappa_4^B(T) \hat \mu_B^4 +\kappa_6^B(T) \hat \mu_B^6+\mathcal{O}(\hat \mu_B^8)\Big)\,.\label{eq:T_res}
\end{align}
Here, $\kappa_2^B(T), \kappa_4^B(T), \kappa_6^B(T)$ are expanding coefficients of different orders.

In \Fig{fig:chi1_muB}, we show the $\chi^B_1/\hat \mu_B$ as a function of temperature with real and imaginary baryon chemical potentials in the range of $|\hat\mu_B|\leq 7\pi/8 $. Obviously, one arrives at $\chi^B_1/\hat \mu_B = \chi^B_2$ at $\hat \mu_B=0$. As we can see, for a fixed value of $\hat \mu_B$, either real or imaginary chemical potential, $\chi^B_1/\hat \mu_B$ always increase with the temperature monotonically, which is a necessary condition for the rescaling relation in \Eq{eq:chi1_res}. For a fixed temperature, the ratio $\chi^B_1/\hat \mu_B$ also increases with $\hat\mu_B^2$ monotonously. The calculated results in \Fig{fig:chi1_muB} also indicate that the expansion method is only suitable for the temperature around the phase transition \cite{Borsanyi:2021sxv}, i.e., $T \in [80,220]$ MeV for this work, because the ratio $\chi^B_1/\hat \mu_B$ becomes flat at low or high temperatures. Furthermore, at large $|\hat \mu_B|$, the shape of $\chi^B_1/\hat \mu_B$ becomes different from that of $\chi^B_2(\mu_B=0)$, which implies that the rescaled temperature and expansion scheme may no longer work in that region.

The coefficients $\kappa^B_{2n}$ can be calculated by fitting the results at zero and imaginary chemical potentials. In \Fig{fig:con_ext}, we show the rescaled temperature $T'(T,\hat\mu_B)$, which is defined in \Eq{eq:chi1_res}, as a function of $\hat \mu_B^2$. We choose several original temperatures in the range of  $T\in [108,196]$ MeV. The dots denote the results calculated directly from the fRG approach at both the imaginary and real chemical potentials. The red bands stand for the polynomial fitting of fRG data at zero and imaginary chemical potentials, cf. \Eq{eq:T_res}, which are also extrapolated into the regime of real baryon chemical potentials. More specifically, the bands are determined by fitting the results of $\mathrm{Im} \, \hat \mu_B \in [0,|\hat \mu_B|_{\mathrm{max}}]$. The upper bound $|\hat \mu_B|_{\mathrm{max}}$ is varied in the range of $[2.0,2.4]$ MeV, in order to investigate errors of the fitting, denoted by the width of bands. The comparison between the extrapolation of fitting at imaginary chemical potentials and the direct fRG results at real chemical potentials constitutes a nontrivial test. One can see that the agreement in the region of $|\hat \mu_B^2|\lesssim 6$, that is $|\hat \mu_B|\lesssim 2.5$, is very well, while there is some difference beyond this regime.

The coefficients $\kappa^B_{2n}$ can also be calculated from the Taylor expansion coefficients.
With the Taylor expansion \labelcref{eq:Taylor}, we arrive at
\begin{align}
\frac{\chi_1^B(T,\hat \mu_B)}{\hat \mu_B}=\sum_{n=1}\frac{1}{(2n-1)!}\chi^B_{2n}(T,0) \cdot \hat \mu_B^{2n-2}\,.
\label{eq:Talyorchi1}
\end{align}
Comparing \labelcref{eq:Talyorchi1} with \labelcref{eq:chi1_res} and \labelcref{eq:T_res}, one obtains the relations between the two sets of expansion coefficients
\begin{align}
\frac{\chi_4^B(T)}{3!}=&\frac{\partial \chi_2^B}{ \partial T} T \kappa_2^B(T)\,,\\[2ex]
\frac{\chi_6^B(T)}{5!}=&\frac{\partial \chi_2^B}{ \partial T} T \kappa_4^B(T) +\frac{1}{2!}\frac{\partial^2 \chi_2^B}{(\partial T)^2} T^2 (\kappa_2^B(T))^2\,,\\[2ex]
\frac{\chi_8^B(T)}{7!}=&\frac{\partial \chi_2^B}{ \partial T} T \kappa_6^B(T)
+\frac{1}{2!}\frac{\partial^2 \chi_2^B}{(\partial T)^2} T^2 (2 \kappa_2^B(T)\kappa_4^B(T)) \nonumber \\[1ex]
&+\frac{1}{3!}\frac{\partial^3 \chi_2^B}{(\partial T)^3} T^3 (\kappa_2^B(T))^3\,.
\end{align}
Then, the coefficients in turn can be solved order by order:
\begin{align}
\kappa^B_2 =& \frac{\chi^B_4}{6 T {\chi^B_2}'}\,,\label{eq:Talyorcalkappa2}\\[2ex]
\kappa^B_4 =&\frac{3 ({\chi^B_2}')^2 \chi^B_6-5 {\chi^B_2}'' (\chi^B_4)^2}{360 T ({\chi^B_2}')^3}\,,\label{eq:Talyorcalkappa4}\\[2ex]
\kappa^B_6 =&\Big[105 ({\chi^B_2}'')^2 (\chi^B_4)^3-63 {\chi^B_2}''({\chi^B_2}')^2 \chi^B_4  \chi^B_6 -35 {\chi^B_2}''' \nonumber \\[1ex]
&\times{\chi^B_2}' (\chi^B_4)^3+9 ({\chi^B_2}')^4 \chi^B_8\Big] \Big[45360 ({\chi^B_2}')^5 T\Big]^{-1}\,.
\label{eq:Talyorcalkappa6}
\end{align}
Here the prime, e.g. ${\chi^B_2}'$, denotes the derivative with respect to the temperature. Obviously, in the new expansion scheme, the coefficients $\kappa^B_{2n}$ encode the information of $\chi^B_{2} \cdots \chi^B_{2n+2}$ as well as $1\cdots n$-th order temperature derivatives of ${\chi^B_2}$ at vanishing chemical potential. For lattice QCD, it is a numerical challenge to precisely calculate the  $n$-th order temperature derivatives of ${\chi^B_2}$. On the other hand, it is feasible for lattice QCD to obtain the coefficients $\kappa^B_{2n}$ by fitting ${\chi^B_2}$ at several imaginary chemical potentials with \Eq{eq:T_res} and calculate the temperature derivatives of ${\chi^B_2}$ in turn \cite{Borsanyi:2021sxv}. 

In \Fig{fig:kappa24}, the coefficients calculated from the fitting of imaginary chemical potentials and the Taylor expansion coefficients in \labelcref{eq:Talyorcalkappa2,eq:Talyorcalkappa4,eq:Talyorcalkappa6} are presented. Obviously, the results obtained from two different methods agree with each other very well. Our results are also comparable with the lattice results \cite{Borsanyi:2021sxv}, in the range of $0.015 \lesssim \kappa^B_2 \lesssim 0.03$ with a slight increase in the regime of high temperature, while the high-order coefficients $\kappa^B_4$ amd $\kappa^B_6$ are very close to zero. In the following, we use the  coefficients calculated from the Taylor expansion coefficients and ignore the errors.

In this expansion scheme, the pressure reads:
\allowdisplaybreaks{
\begin{align}
\frac{p(T,\hat \mu_B)-p(T,0)}{T^4}=&\int_{0}^{\hat\mu_B} d \hat \mu_B' \chi^B_1(T,\mu_B')\nonumber\\[2ex]
=&\int_{0}^{\hat\mu_B} d \hat \mu_B' \hat \mu_B' \chi^B_2(T',0)\,.
\end{align}
The first three order generalized susceptibilities are given as:
\begin{align}\label{eq:chi_cal}
\chi^B_1(T,\mu_B)=&\hat \mu_B \chi^B_2(T',0)\,,\\[2ex]
\chi^B_2(T,\mu_B)=&\chi^B_2(T',0) + \hat \mu_B \frac{\partial \chi^B_2(T',0)}{\partial T'} \frac{\partial T'}{\partial \hat \mu_B}\,,\\[2ex]
\chi^B_3(T,\mu_B)=& 2 \frac{\partial \chi^B_2(T',0)}{\partial T'} \frac{\partial T'}{\partial \hat \mu_B}\nonumber \\[1ex]
&+\hat \mu_B \Bigg (
\frac{\partial^2\chi^B_2}{{\partial T'}^2} \bigg(\frac{\partial T'}{\partial \hat \mu_B}\bigg)^2 +
\frac{\partial \chi^B_2}{\partial T'}\frac{\partial^2 T'}{{\partial \hat \mu_B}^2} \Bigg )
\,.
\end{align}
}

\section{numerical results}
\label{sec:res}

%
\begin{figure*}[t]
\includegraphics[width=1\textwidth]{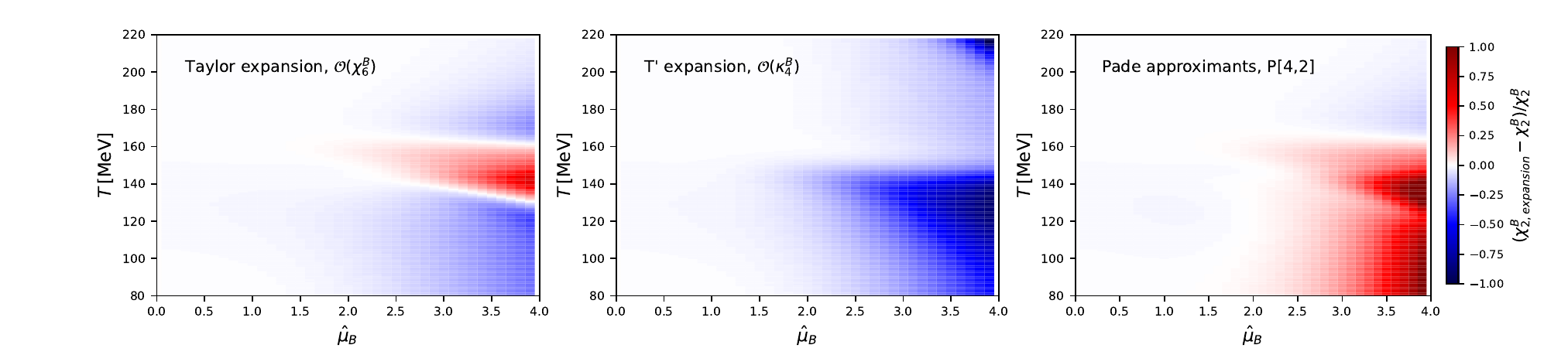}
\includegraphics[width=1\textwidth]{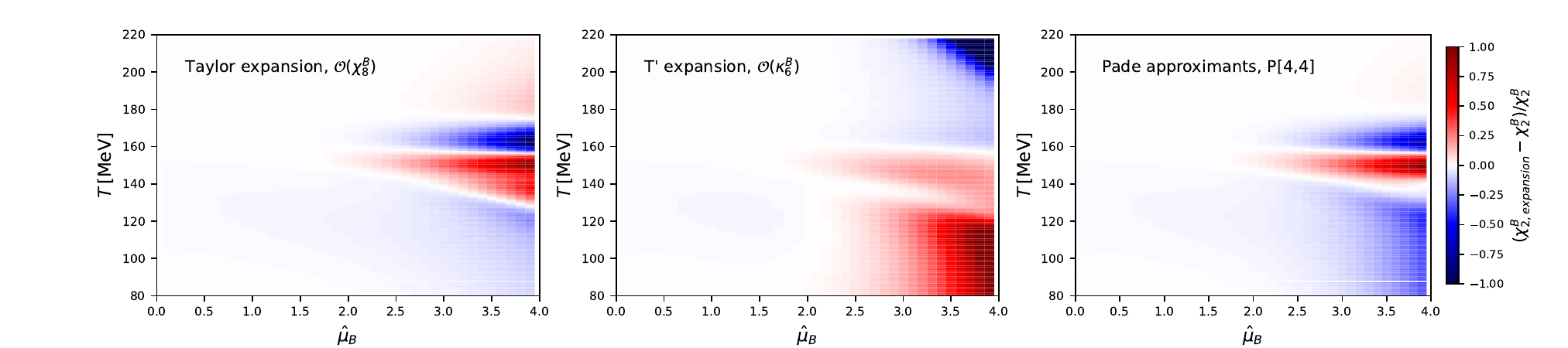}
\caption{Relative errors of different expansion schemes for $\chi_2^B$ in the plane of $T$ and $\hat \mu_B$.}\label{fig:heatchi}
\end{figure*}
%

%
\begin{figure}[t]
\includegraphics[width=0.5\textwidth]{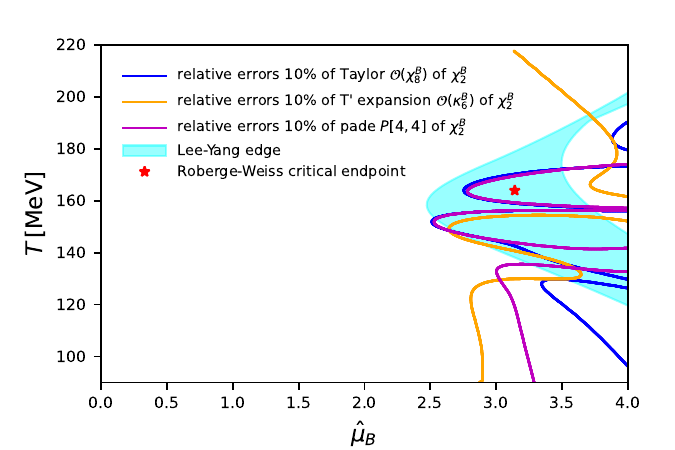}
\caption{Lee-Yang edge singularities and relative errors of different expansion schemes for $\chi_2^B$ in the plane of $T$ and $\hat \mu_B$.}\label{fig:Rconv}
\end{figure}
%

We plot the pressure and the first three order generalized susceptibilities at several temperatures around the critical temperature in \Fig{fig:pre_sus}, which are calculated from the Taylor expansion, the Pad\'e approximants, as well as the $T'$ expansion. The full results from fRG are also plotted for comparison. In order to make a meaningful comparison, we only show results using the first eight order generalized susceptibilities, i.e. the Taylor expansion up to 8-th order, the Pad\'e approximant $P[4,4]$ and the $T'$ expansion of $\kappa^B_6$ order. For the pressure, all expansion schemes of different types converge well, and there is only a slight divergence between the full calculated results and the expansion results for $\hat \mu_B\gtrsim 3.5$. For the generalized susceptibilities, the consistent region becomes smaller with the increase of $\mu_B/T$, which is  $\hat \mu_B\lesssim 3.0, 2.0, 1.5$ for $\chi_1^B,\chi_2^B,\chi_3^B$ respectively.

In order to investigate the errors of different expansion schemes in detail, we show the relative difference of $\chi_2^B$ between the expansion results and full results in the plane of $T$ and $\hat \mu_B$ in \Fig{fig:heatchi}. Subplots in the first line denote the results obtained with the expansion up to the order of $\mathcal{O}(\chi_6^B)$, and those in the second line to $\mathcal{O}(\chi_8^B)$. The consistent region of the second line is a bit larger than that in the first line, which implies that the convergence could be improved mildly by including higher-order susceptibilities. For the Taylor expansion, one can see an alternant structure at large $\hat \mu_B$, that is related with the oscillatory structure of the highest order susceptibility. The results of $T'$ expansion of order $\mathcal{O}(\kappa_4)$ are always smaller than the full results, while the structure becomes a bit complicated for the order $\mathcal{O}(\kappa_6)$. The convergence radius near the critical temperature is smaller than that in the regimes of high or low temperature. Roughly speaking, the convergence radius is about $\hat \mu_B \sim 2$ for $\chi_2^B$ in the vicinity of chiral phase transition. A simple explanation is as follows. At low temperature, the system tends towards hadron resonance gas, i.e. $\chi_{2n}^B/\chi_2^B=1$ for $n\geq1$, and at high temperature, it tends towards the Stefan-Boltzmann limit, i.e. $\chi_{2n}^B/\chi_2^B=0$ for $n\geq3$. Both cases give rise to an infinite convergence radius. Whereas in the proximity of the chiral phase transition, the high order susceptibilities oscillate and the convergence radius is a finite value.

Strictly speaking, the convergence radius of an expansion can only be defined when it is expanded to infinite orders. In our case it implies that we have to know the information on the baryon susceptibilities of infinite orders. As we have mentioned above, the convergence radius of the expansion schemes might be constrained by singularities in the complex plane of chemical potential. Here we consider the Lee-Yang edge singularities, whose convergence radius is given from scaling analysis in \cite{Mukherjee:2019eou}:
\begin{align}
R_{\text{conv}}=\bigg |\frac{z_c}{z_0} \bigg(\frac{m_l^{\text{phys}}}{m_s^{\text{phys}}}\bigg)^\frac{1}{\beta\delta} -\frac{T-T_c^0}{T_c^0}\bigg |^{\frac{1}{2}} \frac{1}{\sqrt{\kappa_2}}.\label{eq:conver-LY}
\end{align}
Here, $m_l^{\text{phys}}/m_s^{\text{phys}}=1/27$, and $\beta,\delta$ are critical exponents, and we use fRG results with LPA truncations in \cite{Chen:2021iuo}, i.e. $\beta= 0.3989$, $\delta=4.975$. The curvature of chiral phase boundary is found to be $\kappa_2=0.0184$ in our low energy effective theory. Note that $\kappa_2$ here is different from the coefficients in \labelcref{eq:T_res}, more details about the curvature of phase boundary can be found, e.g., \cite{Fu:2019hdw, Fu:2021oaw}. The critical temperature in the chiral limit $T_c^0=142.6$ MeV is obtained from fRG calculation in \cite{Braun:2023qak}. The scaling variable $z_c=|z_c|e^{i \frac{\pi}{2 \beta\delta}}$ with $|z_c|=1.665$ and $z_0\in[1,2]$ are suggested in \cite{Mukherjee:2019eou}. With the inputs above, one could estimate the convergence radius of Lee-Yang edge singularities via \labelcref{eq:conver-LY}. The locations of Lee-Yang edge singularities can also be calculated \cite{Mukherjee:2021tyg, Wan:2024xeu}. Moreover, the Roberge-Weiss critical end point is associated with Lee-Yang singularities \cite{Clarke:2023noy, Guenther:2022wcr}, which is located at $\mu_q^{RW}=i \frac{\pi}{3}T$, i.e. $\hat \mu_B^{RW}=i\pi$ \cite{Braun:2009gm, Sun:2018ozp}. The Roberge-Weiss phase transition temperature is found to be $T^{RW}=164$ MeV in our calculations.

In \Fig{fig:Rconv}, we show the lines of relative errors 10\% for the Taylor expansion, the Pad\'e approximants and the $T'$ expansion of $\chi_2^B$ in the plane of $T$ and $\hat \mu_B$, and compare them with the Lee-Yang edge singularities estimated above. The convergence radius of the $T'$ expansion become small at high temperatures has been explained in \cref{subsec:T_exp}. It is found that all expansion schemes have almost the same convergence radius around the critical temperature, and they consist with the Lee-Yang edge singularities. The Roberge-Weiss phase transition singularity is also located within this region.


\section{Conclusion}
\label{sec:sum}

In this work, the convergence of different expansion schemes at finite baryon chemical potentials, including the conventional Taylor expansion, the Pad\'e approximants, and the $T'$ expansion proposed recently in lattice QCD simulations, have been investigated in a low energy effective theory within the fRG approach. This is facilitated by the full results of baryon number fluctuations at both real and imaginary chemical potentials, directly calculated in our approach. 

We employ two different methods to calculate the expanding coefficients of $T'$ expansion, i.e., fitting the $T'$ at imaginary chemical potentials or using relations between the expanding coefficients of $T'$ expansion and those of Taylor expansion. These two methods provide us with consistent results, which are also comparable to lattice simulations.

The pressure is obtained within the three different expansion schemes all up to the expanding order $\mathcal{O}(\mu_B^8)$, from which we calculate the baryon number fluctuations of first three orders, i.e., $\chi_1^B$, $\chi_2^B$, $\chi_3^B$. The pressure and the fluctuations of first three orders are found to be consistent with the full results within the regions $\mu_B/T \lesssim$ 3.5, 3.0, 2.0, 1.5, respectively. The consistent region near the critical temperature is smaller than those at high or low temperature. We also compare the results obtained with the expansion order up to $\mathcal{O}(\mu_B^6)$ and those up to $\mathcal{O}(\mu_B^8)$, which indicates that it could enlarge a bit the consistent region by including higher-order expansions. It is found that the $T'$ expansion or the Pad\'e approximants would hardly improve the convergence of expansion in comparison to the conventional Taylor expansion, within the expansion orders considered in this work. Furthermore, We also estimate the singularities in the complex plane of chemical potential, arising from the Lee-Yang edge singularities and Roberge-Weiss phase transition. The consistent regions of the three different expansions are in agreement with the convergence radius of the Lee-Yang edge singularities.


\begin{acknowledgments}
We thank Mei Huang and Yang-yang Tan for their valuable discussions. This work is supported by the National Natural Science Foundation of China under Grant No. 12175030 and Fundamental Research Funds for the Central Universities No. E3E46301X2. 
\end{acknowledgments}



%
%



\bibliography{ref-lib}

\end{document}